\documentclass[a4paper,prl,twocolumn,amsmath,amssymb,longbibliography]{revtex4-1}
\usepackage{bm}
\usepackage{graphicx,color}
\usepackage{color,soul}
\usepackage[colorlinks,linkcolor=magenta,citecolor=magenta]{hyperref}

\newcommand{\mean}[1]{\langle #1 \rangle}

\renewcommand{\vec}[1]{\mathbf #1}

\newcommand{\al}{\alpha}

\newcommand{\sig}{\sigma}

\newcommand{\x}{\vec r}

\newcommand{\im}{\text{i}}

\AtBeginDocument{%
	\newwrite\bibnotes
	\def\bibnotesext{Notes.bib}
	\immediate\openout\bibnotes=\jobname\bibnotesext
	\immediate\write\bibnotes{@CONTROL{REVTEX41Control}}
	\immediate\write\bibnotes{@CONTROL{%
			apsrev41Control,author="08",editor="1",pages="1",title="0",year="1"}}
	\if@filesw
	\immediate\write\@auxout{\string\citation{apsrev41Control}}%
	\fi
}%

\begin{document}

\title{In pursuit of the tetratic phase in hard rectangles}

\author{Denis Dertli}
\author{Thomas Speck}
\affiliation{Institute for Theoretical Physics IV, University of Stuttgart, Heisenbergstr. 3, 70569 Stuttgart, Germany}

\begin{abstract}
	We numerically investigate two-dimensional systems of hard rectangles at constant pressure through extensive hard-particle Monte Carlo simulations. We determine the complete phase diagram as a function of packing fraction and aspect ratio, which consists of four distinct phases. At very high packing fractions, particles form a smectic solid for all aspect ratios. Rod-like particles with large aspect ratio assemble in an intervening nematic phase, which is displaced by a ``tetratic'' phase (also called biaxial nematic) for moderately elongated rectangles. Surprisingly, we find evidence that the transition from tetratic to smectic is weakly discontinuous at variance with previously proposed two-step scenarios for the melting of hard particles.
\end{abstract}

\maketitle


The emergence of order in microscopic assemblies of hard bodies driven entirely through entropy is a fascinating phenomenon defying the wide-spread notion that associates entropy with chaos and disorder. The prediction that hard spheres order into a regular crystal structure is one of the early triumphs of computational statistical mechanics~\cite{alder57,wood57}, which was confirmed by the ground breaking experiments of Pusey and van Megen~\cite{pusey86}, for a comprehensive review see also~\cite{royall23}. Even earlier, Onsager predicted that hard elongated rods can develop orientational order while positions remain disordered, which highlights the importance of shape~\cite{onsager49}. More recent work demonstrates the wide variety of structural and dynamic phases that are unlocked through varying particle shape~\cite{bolhuis97,haji-akbari11,damasceno12,ni12,haji-akbari15,anderson17}.

Another crucial factor is dimensionality~\cite{mermin66,hohenberg67}, with two-dimensional materials displaying a number of peculiar properties when compared to their bulk behavior. Indeed, already for hard discs is the phase diagram more involved compared to three dimensions with a hexatic phase intervening between fluid and solid~\cite{strandburg88,thorneywork17}. The precise nature of the phase transitions has been a long-standing matter of debate until settled through extensive numerical simulations to be discontinuous (fluid-hexatic) and continuous (hexatic-solid)~\cite{bernard11,engel13,kapfer15}. Anderson and coworkers have performed a comprehensive numerical investigation for regular $n$-gons~\cite{anderson17}. Beside the disk-like scenario for large $n\geqslant 7$ they also find a discontinuous transition from fluid to solid without intervening phase (for hard pentagons with $n=5$, see also~\cite{schilling05,zhao09}) and two continuous transitions (corresponding to the original scenario of the Kosterlitz-Thouless-Halperin-Nelson-Young theory~\cite{kosterlitz73,halperin78}) for triangles, squares, and hexagons. Experimentally, granular~\cite{walsh16} and colloidal particles with complex shapes~\cite{hou20} have been studied.

\begin{figure}[b!]
	\centering
	\includegraphics{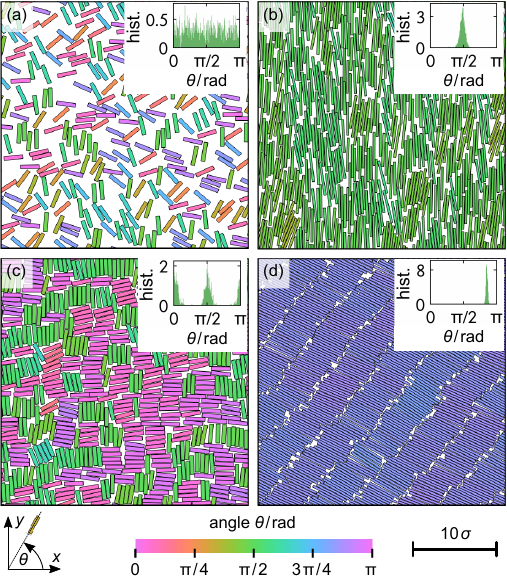}
	\caption{Simulation snapshots for the four observed phases: (a)~isotropic (I) with $\alpha=5$, $p=1.3$, $\phi\simeq0.347$, (b)~nematic (N) with $\alpha=15$, $p=5.0$, $\phi\simeq 0.596$, (c)~tetratic (T) with $\alpha=5$, $p=8.8$, $\phi\simeq0.730$, and (d)~smectic (S) with $\alpha=15$, $p=12.8$, $\phi\simeq0.812$. The insets show the distribution of the angle $\theta$ enclosed with the $x$-axis. Notably, the tetratic phase (c) exhibits two equal peaks, whereas nematic (b) and smectic (d) phase show the same orientational order and are distinguished by the breaking of translational symmetry in the smectic phase.}
	\label{fig:snap}
\end{figure}

The arguably simplest shape in two dimensions is the hard rectangle. While the limits of (rounded) squares~\cite{zhao11,avendano12,anderson17} and elongated rods~\cite{onsager49,bates00,vink09} have been studied extensively, intermediate aspect ratios $\al$ have received far less attention. Theoretical phase diagrams (based on scaled particle theory) have been constructed~\cite{martinez-raton05,martinez-raton06} but, somewhat surprisingly, there is no comprehensive numerical phase diagram. Here we fill this gap and provide a complete phase diagram as a function of packing fraction and particle aspect ratio featuring four phases that are illustrated in Fig.~\ref{fig:snap}. Intriguingly, we see a clear delineation of a genuine ``tetratic'' phase [Fig.~\ref{fig:snap}(c)] characterized by a distribution of orientations that shows two peaks. For squares ($\al=1$), Wojciechowski and Frenkel hypothesized that such a phase with four-fold neighbor symmetry would take the place of the more familiar hexatic phase for discs (which have six neighbors)~\cite{wojciechowski04}. Donev \emph{et al.} have performed simulations of ``dominos'' ($\al=2$) and found clear evidence for tetratic order between fluid and solid~\cite{donev06}. In earlier studies of ``discorectangular'' particles a tetratic phase was not detected~\cite{bates00}. Experimentally, effective hard rectangles have been realized as standing discs on a substrate~\cite{zhao07}, for which tetratic order has been reported. Interest in the precise phase behavior of hard particles in two dimensions is rekindled by the recent progress on the self-assembly of DNA-based tiles~\cite{khmelinskaia21,franquelim21,zhan23}. Grafted to lipid membranes and endowed with local and specific interactions through DNA hybridization, they open new avenues to functional synthetic nanostructures.


We study $N$ identical rectangular particles with aspect ratio $\al$ and fixed area $\sig^2$ at constant pressure $p$ moving in two dimensions with periodic boundary conditions~\cite{frenkel23}. If not stated otherwise, $N\approx1300$ depending on the box aspect ratio to guarantee a defect-free initial configuration. Particles only interact through their excluded volume and we perform extensive hard-particle Monte Carlo simulations utilizing \texttt{HOOMD-blue}~\cite{anderson08,glaser15,anderson16,anderson20}. Throughout, we employ dimensionless quantities with length unit $\sig$ and energies in units of the thermal energy $k_\text{B}T$. A single configuration thus comprises the centroid positions $\x_k$ and the angles $0\leqslant\theta_k<\pi$ enclosed with the $x$-axis for each particle. Three protocols are implemented distinguished by their initial state: (i)~$\mathcal I_+$ starts from a disordered state and then the pressure is increased until the target pressure is reached while (ii)~$\mathcal S_-$ starts from a perfect smectic solid and the pressure is decreased. (iii)~In addition, for $\al=5$ we prepare a perfect ``checkerboard'' within the tetratic phase (at $\phi \simeq 0.72$) as initial configuration (protocol $\mathcal T_\pm$). The checkerboard unit cell is composed of $4 \times 5$ rectangles as follows: Always five particles form a small group where all particles point in the same direction, while the nearest neighbor groups are aligned in the perpendicular direction. All protocols prepare configurations at a target pressure through compressing/expanding a protocol-dependent initial configuration. This is followed by a relaxation run of $\tau_\text{r}$ Monte-Carlo time steps, after which we collect configurations for analysis (see supplemental information~\cite{sm} for further details).

To characterize single configurations harvested from the simulations, we employ three order parameters. The first two are
\begin{equation}
	\hat\psi_\nu = \left|\frac{1}{N}\sum_{k=1}^N e^{\im\nu\theta_k}\right|^2
	\label{eq:psi:k}
\end{equation}
with $\nu=2,4$ and $0\leqslant\hat\psi_\nu\leqslant1$. These measure orientational order with $\hat\psi_\nu\simeq 0$ if the distribution of orientations is uniform. Both the nematic and smectic phase are characterized by a distribution of orientations exhibiting a single peak corresponding to the preferred orientation, the unit-length director. In the smectic phase [Fig.~\ref{fig:snap}(d)], the distribution is much narrower than in the nematic phase and smectic layers form separated by a small gap. Importantly, for two peaks $\hat\psi_4$ becomes non-zero while $\hat\psi_2\simeq 0$, whereas for a single peak both order parameters are non-zero. Through the third order parameter $\psi_x$, we measure the breaking of translational symmetry in the smectic phase (details on its construction can be found in the supplemental information~\cite{sm}). In addition, we record the actual area $\hat A$ for each configuration, from which we determine the packing fraction $\phi=N/\mean{\hat A}$. While we control the pressure $p$ in simulations, in the following we will mostly report $\phi$ as the control parameter. To extract the phase behavior, we calculate the expectation values $\psi_\nu=\mean{\hat\psi_\nu}$ of the order parameters and their susceptibilities
\begin{equation}
	\chi_\nu = \mean{\hat\psi_\nu^2} - \mean{\hat\psi_\nu}^2,
\end{equation}
where the brackets $\mean{\cdot}$ denote the equilibrium expectation value. To improve the statistical efficiency and to interpolate between pressures, we employ the \emph{multistate Bennett acceptance ratio} method (MBAR)~\cite{shirts08,shirts20}. In the following, we identify transitions through peaks of the susceptibilities.


\begin{figure}[b!]
	\centering
	\includegraphics{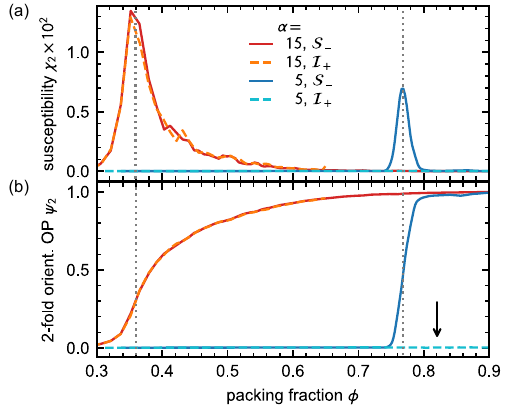}
	\caption{Orientational order measured through Eq.~\eqref{eq:psi:k} with $\nu=2$ for high ($\al=15$) and low ($\al=5$) aspect ratios from expansion ($\mathcal S_-$, solid lines) and compression ($\mathcal I_+$, dashed lines) protocols. (a)~Susceptibility $\chi_2$. Peaks that we identify with transitions are indicated by gray dashed lines: continuous isotropic-nematic (IN) for $\al=15$ at $\phi^\ast\simeq0.36$ and discontinuous tetratic-smectic (TS) for $\al=5$ at $\phi^\ast\simeq0.77$. See discussion in main text. (b)~Order parameter $\psi_2$. Note the strong hysteresis for $\al=5$, where no signal is found along the expansion route (arrow).}
	\label{fig:op2}
\end{figure}

Figure~\ref{fig:op2}(a) shows the susceptibility $\chi_2$ as a function of packing fraction $\phi$ for hard rectangles with $\al=5$ and $\al=15$. We see that there is a clear peak for both aspect ratios indicating a transition towards orientational order [Fig.~\ref{fig:snap}(b)]. However, the nature of the transition is different in both cases. For strongly elongated rectangles ($\al=15$), the peak of $\chi_2$ occurs at low packing fraction $\phi^\ast\simeq0.36$ and $\psi_2$ [Fig.~\ref{fig:op2}(b)] exhibits a smooth increase for both protocols $\mathcal I_+$ and $\mathcal S_-$. At these packing fractions, the translational order parameter $\psi_x$ does not show any signal~\cite{sm}. This continuous transition thus occurs from the isotropic phase to the nematic phase (IN). At a larger packing fraction also the translational order parameter rises (its susceptibility peaks at $\phi^\ast\simeq0.73$~\cite{sm}) indicating the transition to the smectic phase [Fig.~\ref{fig:snap}(d)]. Across this second transition, the nematic order parameter $\psi_2\simeq 1$ remains a smooth function.

\begin{figure}[b!]
	\centering
	\includegraphics{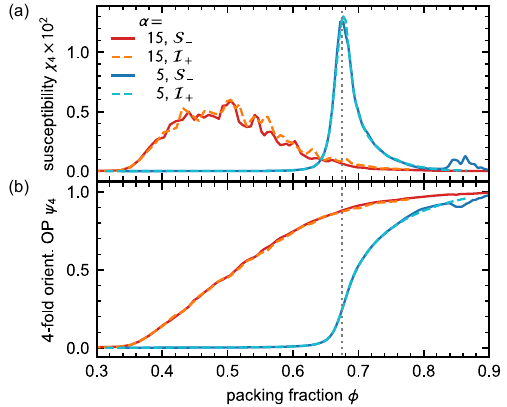}
	\caption{Analogous to Fig.~\ref{fig:op2} but plotting (a)~$\chi_4$ and (b)~$\psi_4$. The peak of $\chi_4$ (gray dashed line) corresponds to the isotropic-tetratic (IT) transition. The very broad peak of $\chi_4$ for $\al=15$ lies within the nematic phase and is not identified as a transition.}
	\label{fig:op4}
\end{figure}

In contrast, for small aspect ratio $\al=5$ the order parameter $\psi_2$ in Fig.~\ref{fig:op2}(b) displays a sigmoidal profile as expected for a discontinuous transition. Further evidence comes from the strong hysteresis between compression ($\mathcal S_-$) and expansion ($\mathcal I_+$) route, where for the latter $\psi_2\simeq 0$ shows no signs of ordering. Importantly, the jump of $\psi_2$ coincides with the jump of the translational order parameter $\psi_x$~\cite{sm}. We thus conclude that this jump corresponds to a transition into the smectic phase. From the behavior of $\psi_2$ alone one might be tempted to identify the other phase as isotropic since $\psi_2$ exhibits a single jump. However, turning to $\chi_4$ in Fig.~\ref{fig:op4}(a) reveals another narrow peak at a smaller packing fraction $\phi^\ast\simeq0.68$ before entering the smectic phase. Configurations for $\phi>0.68$ [Fig.~\ref{fig:snap}(c)] exhibit short-range translational order together with orientational ordering along two perpendicular axes. Consequently, the distributions of orientations show two approximately equal populations separated by $\pi/2$. We thus associate this peak with the transition between the isotropic and a genuine \emph{tetratic} phase. Across the tetratic-smectic (TS) transition, now $\psi_4$ remains smooth.

We note that $\psi_4$ also increases for $\al=15$ [Fig.~\ref{fig:op4}(b)], but with a rather gentle slope and a corresponding wide peak of the susceptibility $\chi_4$ [Fig.~\ref{fig:op4}(a)]. Crucially, $\psi_4$ alone is not sufficient to distinguish tetratic order since it becomes non-zero also for nematic order. Only the combination of non-vanishing $\psi_4$ together with $\psi_2\simeq0$ is indicative of tetratic order characterized by a bimodal angle distribution. We thus conclude that there is no tetratic phase anymore for $\al=15$, and that the order for the transition to the smectic phase has changed from discontinuous for small aspect ratios $\al$ to continuous for large $\al$. Before addressing intermediate aspect ratios, we turn to the nature of the transition into the smectic phase for small aspect ratios.

\begin{figure}[t]
	\centering
	\includegraphics[width=\linewidth]{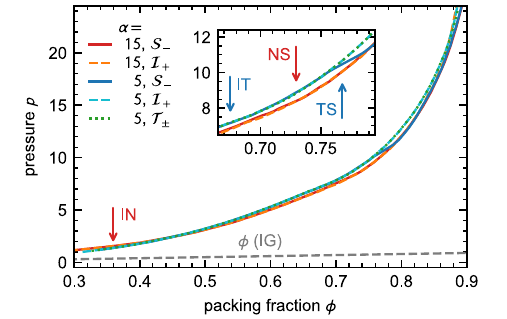}
	\caption{Equation of state. Pressure $p(\phi)$ plotted as function of the average packing fraction $\phi$. The gray dashed line shows the ideal gas pressure. The inset zooms into the range of packing fractions where phase transitions occur (arrows).}
	\label{fig:eos}
\end{figure}

Figure~\ref{fig:eos} shows the equation of state $p(\phi)$ (note that the pressure is the control variable and $\phi$ is determined from the average area), again for $\al=5$ and $\al=15$ below and above the onset of the tetratic phase. As expected, no signature of the continuous transitions is seen in these curves. This is different for the TS transition for $\al=5$, where we confirm the hysteresis in the equation of state: The expansion route $\mathcal S_-$ switches from the smectic branch to the tetratic branch, whereas the compression route $\mathcal I_+$ continues along the (now metastable) tetratic branch, corroborating our identification of this transition as discontinuous. We note, however, that the crossing between smectic and tetratic branch is rather gradual, which we attribute to our simulations being performed in the constant-pressure ensemble. Moreover, a closer look reveals that in trajectories sampled in the neighborhood of the TS transition a relaxation time of $\tau_\text{r} \sim 10^7$ is insufficient to reach the stationary regime and the two-fold orientation order parameter $\psi_2$ (and also $\psi_x$) decreases further as $\tau_\text{r}$ is increased (see supplemental information~\cite{sm} for a more detailed discussion on finite-time and finite-size effects).

\begin{figure}[t]
	\centering
	\includegraphics[width=\linewidth]{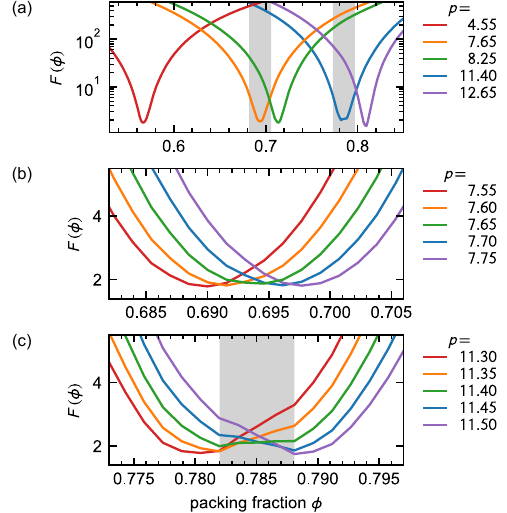}
	\caption{Free-energy profiles $F(\phi)$ of the packing $\phi$ for $\al = 5$ and protocol $\mathcal S_-$. (a)~Selected pressures covering a wide range of $\phi$. We zoom into the two shaded areas: (b)~Across the IT transition ($p^\ast\simeq 7.65$, green), the position of the minimum of $F(\phi)$ evolves continuously from a high to a low value of $\phi$. (b)~Across the TS transition ($p^\ast\simeq 11.40$, green), we now observe a signature of a bimodality within a narrow pressure range corroborating a discontinuous transition. The shaded area indicates the coexistence region.}
	\label{fig:fepphi}
\end{figure}

To gain further insight into the nature of the tetratic-smectic (NS) transition at $\alpha = 5$, we employ MBAR to calculate the probability $P(A)$ of the fluctuating area, from which we determine the free energy $F(\phi)=-\ln P(\phi)$ of the packing fraction (see supplemental information~\cite{sm}). Note that for these runs, the relaxation time $\tau_\text{r} \sim 10^8$ is one order of magnitude longer. Figure~\ref{fig:fepphi}(a) shows free-energy profiles as we increase the pressure $p$ across the first transition from isotropic to tetratic. Distributions are clearly unimodal and their minimum smoothly varies with pressure as one would expect for a continuous phase transition. This changes for higher pressures shown in Fig.~\ref{fig:fepphi}(b), where we see signs for a non-convex distribution. Around the transition pressure $p^\ast\simeq 11.40$, the minimum of $F(\phi)$ jumps. These features further corroborate our identification of the TS transition as discontinuous. The coexistence region is very narrow, $\Delta\phi\simeq 0.0055$. 

While we have presented numerical evidence for the TS transition to be discontinuous, one might still wonder what stabilizes the tetratic over the smectic phase. The tetratic phase is clearly characterized by a bimodal distribution of orientations, but it also shares positional order with the smectic phase. Instead of perfect layers, we find a network of domains that are orientated perpendicular. Blocks of $\al$ particles can flip collectively and form a lattice. Of course, perfect periodic order in two dimensions is suppressed by long-wavelength fluctuations (Goldstone modes)~\cite{mermin66,hohenberg67}. Moreover, for layer dislocations in the smectic phase that cost a finite free energy there is a density $c_\text{D}$ of these defects with a corresponding length $\xi_\text{D}\sim c_\text{D}^{-1/2}$ (certainly larger than the system sizes studied here) beyond which the smectic phase fractures into domains~\cite{toner81}. Treating the system as blocks (super-particles) of $\al$ rectangles, the degeneracy of the two perpendicular orientations raise the entropy of the tetratic phase by roughly $(1/\al)\ln 2$ per particle~\cite{donev06} (whereas the smectic phase is denser). This entropy gain diminishes as the aspect ratio increases.

\begin{figure}[t]
	\centering
	\includegraphics[width=\linewidth]{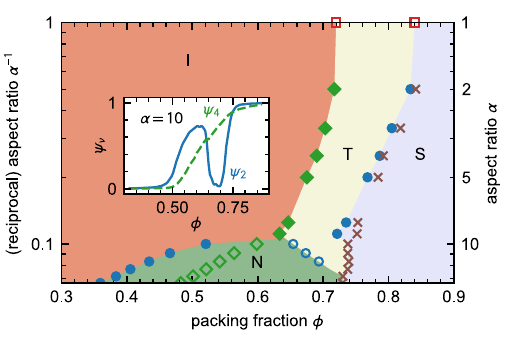}
	\caption{The phase diagram of hard rectangles. Closed symbols indicate the location of peaks in the susceptibilities: $\chi_2$ ($\bullet$), $\chi_4$ ($\blacklozenge$), and $\chi_x$ ($\times$). The open blue symbols correspond to the second peak of $\chi_2$ while the open green symbols indicate the position of the wide peak of $\psi_4$ inside the nematic phase, cf. Fig.~\ref{fig:op4}. Also shown are the phase boundaries for squares ($\al=1$, $\square$) taken from Ref.~\cite{anderson17}. The inset shows $\psi_{2,4}$ for $\al=10$.}
	\label{fig:phasediag}
\end{figure}

Putting everything together, Fig.~\ref{fig:phasediag} shows the resulting phase diagram for hard rectangles as a function of average packing fraction $\phi$ and aspect ratio $\al$. Our numerical evidence suggests the following orders: IT, IN, and NS are classified as continuous, while TS and TN exhibit a discontinuous character. For large aspect ratios the nematic phase occupies a wide range of packing fractions but narrows and finally terminates at $\al\simeq 9$. In the inset of Fig.~\ref{fig:phasediag}, we show the orientational order $\psi_2$ at $\al=10$. We see that $\psi_2$ rises as the system enters the nematic phase ($\psi_4\simeq0$) but then drops to zero before rising again but this time $\psi_4$ is already close to unity. We thus observe two peaks, the first is associated with the IN transition and the second with the NT transition. For smaller aspect ratios $\al<10$, the second peak vanishes and the position of the first peak jumps to coincide with the emergence of translational order (the TS transition). For larger aspect ratios the second peak moves to larger packing fractions but the drop becomes less pronounced.

For squares (and also discs), the computational strategy used here cannot be followed since a local order parameter cannot be defined. Instead, one has to turn to correlations of bond order to identify the ``$n$-tic'' phase. The underlying degeneracy is lifted for rectangles and orientational order (as measured through $\psi_\nu$) is sufficient to identify the various phases. Still, as shown in Fig.~\ref{fig:phasediag}, the tetratic phase smoothly connects to the phase boundaries determined for squares in Ref.~\cite{anderson17}, although our identification of transition orders is at variance with the proposed two continuous transitions for squares.


To conclude, we have numerically determined the full phase diagram for hard rectangles in two dimensions as a function of packing fraction $\phi$ and small to intermediate aspect ratios $2\leqslant\al\leqslant15$. We identify four phases: isotropic (no orientational and no translational order), nematic (uniaxial orientational order), smectic (uniaxial orientational and translational order), and tetratic (biaxial orientational order). The nematic phase in hard rectangles terminates at a finite aspect ratio $\al\simeq 9$, below which it is replaced by the tetratic phase. The line of susceptibility maxima between nematic and tetratic order meets the phase boundary of the isotropic phase in a special ``tricritical'' point that deserves further investigation beyond the scope of this work. Another interesting question is the fate of the tetratic phase for large aspect ratios, and whether it also terminates or tapers into a very thin stripe between nematic and smectic phases.

The order parameters $\psi_\nu$ we have used are proxies that probe the distribution of orientations and are convenient for efficient sampling in finite systems. The underlying order is that of uniaxial (nematic) vs. biaxial (tetratic) orientational order. Increasing the system size towards the thermodynamic limit, the unit director will slowly vary spatially (a consequence of ``gapless'' Goldstone modes), and global order parameters fail in agreement with Kosterlitz-Thouless-type continuous transitions. The same holds in principle for the transition to the smectic solid, but now the condensation into layers frees a tiny portion of the area per particle ($\Delta\phi\simeq 0.0055$) and the transition becomes discontinuous. We thus find a novel scenario in hard rectangles for such two-step melting: the isotropic-tetratic transition is continuous (vs. the discontinuous isotropic-hexatic transition for hard disks) while the tetratic-smectic transition is discontinuous (vs. continuous hexatic-solid for hard disks).


\begin{acknowledgments}
	Simulations have been performed using the bwHPC infrastructure (cluster HELIX) supported by the state of Baden-Württemberg and the Deutsche Forschungsgemeinschaft (DFG) through grant INST 35/1597-1 FUGG. In addition, we gratefully acknowledge computing time granted by SimTech (University of Stuttgart and DFG grant no. 390740016 -- EXC 2075).
\end{acknowledgments}


%

\end{document}